# Coevolution of Symbiotic Species

Boon Tiong Melvin Leok
MSC 609, Caltech, Pasadena, CA 91126-0609, USA
*mleok@cco.caltech.edu*
(818)395-1709

**Abstract**

*This paper will consider the coevolution of species which are symbiotic in their interaction. In particular, we shall analyse the interaction of squirrels and oak trees, and develop a mathematical framework for determining the coevolutionary equilibrium for consumption and production patterns.*

## Contents

# 1 Introduction

Survival strategies in nature and financial interactions often arise in which immediate gain is sacrificed in order to increase the probability of survival in the event of unfavourable conditions.

We shall attempt in this paper to model the evolution of "hedging" and determine the optimal level of hedging in the case of the interaction between squirrels and oak trees. In addition, we shall consider the effect coevolution has on the individual strategies of symbiotic species.

# 2 Evolutionary responses

Species genetically evolve in response to their interaction between the environment and other species.

Intuitively, "hedging"as manifested in squirrels could have arisen as a natural response to the inherent variability of food supplies. Such reserves provide a buffer against periods of scarcity of food, and generally increases the long term viability of squirrels with such a genetic trait. Uncollected reserves eventually grow into new oak trees, thereby increasing the food supply on larger time scales.

## 2.1 Relative contribution of selection pressures

It is unlikely however that the mechanism of misplaced acorns increasing the long-term supply of future acorns would be the major contributing factor in genetically selecting for squirrels with the appropriate level of hedging to predominate. This selection pressure is undifferentiated insofar as the beneficiaries of the behaviour is concerned, and as such cannot exert as substantial a selection pressure as a mechanism such as the seasonal variability of food supply which would differentially benefit individuals with distinct responses.

### 2.2 Complexity of the Genetic Interaction Model

If we naively attempt to model the time dependent foraging intensity which can be considered to be uniquely defined by the genotype *without* any simplifying approximations, we will obtain a model which is computational prohibitive.

If we were to assume that the intensity for each month is discretised into $n$ levels, the number of distinct alleles would be of the order $n^{12}$. The number of allele pairs will be $O(n^{24})$. Thus we would have to evaluate $O(n^{24})$ distinct species per time step. Moreover, if we used a pairwise species interaction model, the order of interactions which have to be considered per time step would be $O(n^{48})$.

We thus see that the time complexity of the naively realistic model makes its analysis intractable. It it thus fruitful to consider reduction methods which reduce the time complexity while retaining the salient aspects of the biological system.

## 3 Overview of Modelling Strategy

This model will represent the dynamics of the phenotypes of squirrels using discrete time, with the month being the unit of consideration. Genetic success, will be decomposed into two components, that associated with zygotic selection and the other with fecundity.

### 3.1 Zygotic Selection

The first component is concerned with the differential mortality of each generation, and the individual alleles are only considered to interact with the environment and food supply for this system of equations. This corresponds to the biological notion of zygotic selection which is associated with the viability of a particular system of behavioural patterns. A behavioural pattern which ensures that there is an adequate amount of food to sustain a population irrespective of variations in food supply would fare well in this component.

### 3.2 Fecundity Selection

The second component will determine the level of fertility of one genotype as compared to another at the end of the year, and serves to model the adaptive advantage of a genotype from the perspective of genetic representation in the next generation. This corresponds to the notion of fecundity selection, and is intimately related to the level of reserves which a population has accumulated over the year.

## 4 Reduction Assumptions

The following reduction assumptions allow us to probe the allele space to a reasonable level of precision without resorting to impractically complex models.

### 4.1 Simplification of the Genetic Model

For the purpose of this discussion, we shall adopt a highly simplified genetic model, in which a single allele represents the genotype of the individual. This is not truly representative of biological reality, in which each genotype is specified by two alleles, but this approximation suffices if we consider the phenotype expression to be dependent only on the average of the behaviours prescribed by the two alleles in the genetic configuration.

The approximation introduced would remain reasonably valid insofar as we do not consider the possibility that certain alleles are dominant, a situation which is excluded by the averaging method we utilise as a mechanism for gene expression. This results in a one-to-one correspondence between phenotypes and genotypes.

As we are only truly concerned with the distribution of phenotypes within the population, the approximation is a reasonable method of reducing the complexity of the problem.

### 4.2 Composite Allele approximation to Allele Pairs

For the purpose of the simulation, we shall coalesce allele pairs of the form $A(q_j)$, $A(2p_i - q_j)$ into a composite allele $A'(p_i)$, since $p_i$ is the average of the intensities $q_j$ and $2p_i - q_j$. Thus, the representation of $A'(p_i)$ is related to the representation of the $A(q_j)$, $A(2p_i - q_j)$ pairs by the following expression,

$$|A'(p_i)| = \sum_j |A(q_j), A(2p_i - q_j)| \tag{1}$$

Notice however that representation in the future generation is dependent upon the adaptive advantage that a particular phenotype provides, which is only dependent on $p_i$. As such the observed distribution of phenotypes by considering only the interaction of the composite alleles would be the same as the more involved case of allele pairs.

## 5 Model Representation

The following section is concerned with two possible methods of modelling the dynamical system, and their relative merits.

### 5.1 Genetic Algorithm Formulation

An obvious candidate for simulating this dynamical system would be the iterative genetic algorithm formulation, in which a population of genotypes interact with the environment and other individuals, and the most successful individuals have the representation of their genotype increased in the gene pool of the next generation.

In order to model this method in a reasonable fashion, it is as noted above, necessary to reduce the complexity of the genotype space. A method of simplifying the representation of the gene structure, and an iterative method of converging upon the optimal configuration is presented below.

#### 5.1.1 Representation of the Genetic Structure of Squirrels

The collection behaviour of squirrels can be considered to vary in both the average intensity and the temporal distribution of such efforts. For example, it might be worthwhile to collect a greater amount of acorns during the fall in anticipation of winter. We shall model this behaviour genetically by considering each individual to have behaviours which are prescribed by a single allele.

Each allele has to encode both the effort allocated towards food collection, and the manner in which this intensity varies as a function of time. For the purpose of this model, we shall use the month as the discrete unit of time. The hedging intensity function shall have a period of 1 year, and the variability as a function of the month will be somewhat approximated using a power spectrum of a Fourier series, thus it has the form,

$$H(t) = \frac{1}{9}\left(\sum_{i=0}^{2} a_i \cos\frac{2\pi it}{12}\right)^2 \tag{2}$$

To further simplify the analysis, we shall discretise the values of $a_i = \left\{\frac{j}{5}|j = 0..5\right\} \in [0, 1]$. The genetic configuration can be uniquely defined by specifying the 3-tuple $(a_0, a_{1,}a_2)$. Since the values of $a_i$'s are discrete, we can have a total of $6^3 = 216$ possible alleles.

Each possible allele can is some sense be considered to be a distinct species which is competing for the same resources, which essentially requires us to model the interaction of 216 species in determining their subsequent representation in the next generation.

The hedging intensity function defined over the allele space and time can be represented by $H_{ijkt}$, $i, j, k \in [0, 1, ...5]$ which corresponds to the intensity function associated with the 3-tuple $(\frac{i}{5}, \frac{j}{5}, \frac{k}{5})$ at time $t$. Explicitly, it takes the form,

$$H_{ijkt} = \frac{1}{9}\left(\frac{i}{5} + \frac{j}{5}\cos\frac{2\pi t}{12} + \frac{k}{5}\cos\frac{4\pi it}{12}\right)^2 \tag{3}$$

#### 5.1.2 Refinement of Optimal Hedging Estimate

Notice that the resolution of the optimal level of hedging is in some sense limited by the discretisation of the values of $a_i$. The estimate of the optimal level of hedging can be further refined iteratively by repeating the simulation for the allele subspace which is spanned by the individual subranges of $a_i$ which the previous iteration suggests the optimal parameters lie within.

We are implicitly assuming that the function of adaptive advantage is sufficiently smooth that our discretisation over the allele space that it allows us to bracket the global maxima of the adaptive function. This thus serves as the theoretical basis for the iterative process of converging upon the maxima of the adaptive function, which corresponds to the optimal temporal distribution of hedging in our model.

### 5.2 Multidimensional Optimisation Formulation

It might seem that the methodology associated with genetic algorithms would be an obvious choice for simulating this system. Note however that since we are concerned with the interaction between the squirrels and the environment as opposed to a competitive system, the fitness function is independent of interaction terms, thus it is possible to formulate an explicit fitness function by evaluating the combined effect of zygotic selection and fecundity selection over a period of a single year.

The Genetic Algorithm approach is only necessary if the fitness function depends upon interaction terms, which would imply that there isn't an absolute fitness associated with a particular system of characteristics, in that the adaptive value associated with a trait would be dependent upon the composition of the population.

### 5.2.1 Optimisation of fitness function over genetic configuration space

If the fitness function could be explicitly expressed as a function of the genotype, it would be possible to reduce the problem to maximising the fitness function over the genetic configuration space. The question of local maximas arise when the initial configuration state is not dense in the configuration space, however by repeating the $n$-dimensional optimisation procedure using a random subset of the configuration space as initial points, we will be able to determine the expected level of "hedging" which should be observed in reality.

The adaptation of the $n$-dimensional optimisation formulation also allows us to discard the discretisation procedure which would be required in considering the genetic algorithm formulation, since *Powell's Method in Multi-dimensions* [1] allows us to search the genetic configuration space without having to arbitrarily discretising the evaluation points. We are also able to consider the 12-dimensional genotype space without approximating it using the power spectrum of a Fourier series as described above.

By considering this formulation, we can attempt to search the 12-dimensional genetic configuration space which corresponds to the hedging intensity of each of the months, and thereby determine the optimal collection and hedging strategy which will be adopted.

## 5.3 Selection of Modelling Approach

We shall adopt the multidimensional optimisation formulation for the rest of the paper. It might well seem that the species interact, since there is a fixed amount of food which much be shared by the entire population, but the decrease in food supply essentially introduces a scaling constant which uniformly affects all individuals, thus for the purpose of proportionate representation in the next generation, this term cancels out, and the interaction term can be ignored for the purposes of our discussion.

# 6 Fitness Function

The representation of an allele in the next generation is determined by the adaptive value of the trait encoded by the allele. The fitness function, depends upon the variability of the food supply to a far greater extent than the contribution of hedging to the long term supply of acorns, since the adaptive value of such a trait acts equally upon the population as a whole, whereas individual reserves have a more significant adaptive value, particularly in times of scarcity.

The fitness function will encompass both zygotic and fecundity selection over a period of one year. The yearly cycle will be evaluated from the period immediately following the previous reproductive period up to the reproductive period of the year. Thus, the survival characteristics will be evaluated for the first 12 months of the cycle, and the amount of reserves remaining will determine the fecundity of the genotype in question.

## 6.1 Acorn Reserves

We might consider the state of each individual allele to be represented by the number of representatives and the amount of reserves accumulated over the last 6 months. Reserves get converted to potential oak trees according to an exponential decay function. Thus the reserves would be represented by the 6-tuple $R(t) = (r_1, r_2, r_3, r_4, r_5, r_6)$. In times of scarcity, reserves are expended from the least recent reserves, resulting in the state of the reserves at the end of the month, $R'(t) = (r'_1, r'_2, r'_3, r'_4, r'_5, r'_6)$. Incrementing the time by a month will result in the new state of reserves to be represented by $R(t+1) = (r_0, dr'_1, dr'_2, dr'_3, dr'_4, dr'_5)$, where $d = 0.6$ is the decay coefficient. The contribution towards the supply of acorns which grown into oak saplings will then be $\left(\sum_{i=1}^{5}(1-d)r'_i\right) + r'_6$. This monthly contribution will be consolidated on a yearly level and will contribute to an increase in the number of acorn producing oak trees 30 years hence.

## 6.2 Acorn Collection and Retrieval

To determine the manner in which a squirrel would manage its acorn reserves, it is necessary to determine the relative energetics of acorn collection and retrieval. A squirrel obtains a certain amount of energy from the acorns it consumes in the previous month, and a portion of that energy is allocated towards the collection of new acorns in the current month. The acorns are always picked freshly but reserves may be retrieved to sustain the effort of further food collection. We shall consider the number of acorns consumed by a squirrel in a normal month to be denoted by $A_T$.

### 6.2.1 Energetics of Acorn Collection

We can model the difficulty of finding food by considering the amount of energy expended to obtain a single acorn. If the supply of acorns is normalised with 1 corresponding to a normal supply of acorns, we might define the energy required to collect an acorn fresh from an oak tree to be,

$$E_c(t) = \frac{E_0}{F(t)} \tag{4}$$

where $F(t)$ corresponds to the amount of food available in the month $t$, and $E_0$ corresponds to the amount of energy required to collect a single acorn under normal conditions.

### 6.2.2 Energetics of Acorn Retrieval

One might consider the energy associated with retrieving an acorn to be dependent upon the amount of acorns in a reserve. We might reasonably consider the energy required to retrieve an acorn from a reserve which is equivalent in size to a month's supply of acorns to be $\frac{1}{10}$ of the energy required to collect them from the trees. Thus the equation will have the form,

$$E_r(t) = \frac{E_0 \cdot A_T}{10 \cdot \left(\sum_{i=1}^{6} r_i\right)} \tag{5}$$

### 6.2.3 Acorns as a source of energy

It is necessary to quantify the amount of energy derived from a single acorn, each acorn will supply $E_a$ of energy when consumed. Of greater importance will be the ratio of energy obtained when consumed against the energy expended during collection. We shall define the ratio to have the value,

$$\frac{E_a}{E_0} = 1.5 \tag{6}$$

### 6.3 Population Dynamics

We will restrict the analysis to the fitness of a single individual of a particular genotype. At the start of the cycle, which is just at the completion of the previous reproductive cycle, we may assume that the majority of the reserves have been expended towards birth and nurturing of the new generation, with the rest of the reserves serving as the energy source for the current population.

We might consider the squirrel to have a sustenance energy level which corresponds to the amount of energy required to sustain a squirrel over a single month, which we shall denote as $E_s$. Note that the percentage of the original population left at time $t$ is denoted by $p(t)$.

The amount of acorns collected by the remaining squirrel population in a given month, $A_c$ would then be,

$$A_c(t) = \frac{p(t) \cdot E_s}{E_a} (1 + H(t)) \tag{7}$$

Where $H(t)$ denotes the amount of hedging which is encoded for the particular month.

The total amount of energy expended, $E_{total}$ to collect these acorns will be,

$$\begin{aligned} E_{total}(t) &= E_c \cdot A_c(t) \\ &= \frac{E_0}{F(t)} \cdot \frac{p(t) \cdot E_s}{E_a} (1 + H(t)) \\ &= \frac{p(t) \cdot E_s}{1.5 \cdot F(t)} (1 + H(t)) \end{aligned} \tag{8}$$

#### 6.3.1 Utilisation of Reserves

In the event that the sustenance energy level is inadequate to allow the squirrel to collect the required number of acorns,

$$p(t) \cdot E_s < E_{total}(t) = \frac{p(t) \cdot E_s}{1.5 \cdot F(t)} (1 + H(t)) \tag{9}$$

the reserves will be expended to allow the squirrel to obtain the required amount of acorns. Thus, the amount of reserves, $n$ which will be consumed will be given by solving the following equation for $n$,

$$\begin{aligned} n &= \frac{\frac{p(t) \cdot E_s}{1.5 \cdot F(t)} (1 + H(t)) + n \cdot \frac{E_0 \cdot A_T}{10 \cdot \left(\sum_{i=1}^{6} r_i\right)} - p(t) \cdot E_s}{E_a} \\ &= \frac{p(t) \cdot \left(\frac{E_s}{1.5 \cdot F(t)} (1 + H(t)) - E_s\right)}{E_a - \frac{E_0 \cdot A_T}{10 \cdot \left(\sum_{i=1}^{6} r_i\right)}} \end{aligned} \tag{10}$$

### 6.3.2 Inadequate Energy Levels for Collection Intensity

Should the amount of reserves be inadequate to compensate for the inadequate consumption, the total energy used in the collection process for new acorns will be,

$$E_{total}(t) = p(t) \cdot E_s + \left(\sum_{i=1}^{6} r_i\right)\left(E_a - \frac{E_0 \cdot A_T}{10 \cdot \left(\sum_{i=1}^{6} r_i\right)}\right) \tag{11}$$

the energy derived from consumed acorns, $E_{consumed}$ would be,

$$\begin{aligned} E_{consumed}(t) &= \frac{E_a \cdot E_{total}(t)}{E_c \cdot (1 + H(t))} \\ &= \frac{E_a\left(p(t) \cdot E_s + \left(\sum_{i=1}^{6} r_i\right)\left(E_a - \frac{E_0 \cdot A_T}{10 \cdot \left(\sum_{i=1}^{6} r_i\right)}\right)\right)}{E_c \cdot (1 + H(t))} \end{aligned} \tag{12}$$

Thus, in such a situation, only a portion of the population will survive, with the rest lost to starvation. The survival proportion is,

$$\begin{aligned} P_{survival}(t) &= \frac{E_{consumed}(t)}{p(t) \cdot E_s} \\ &= \frac{E_a\left(p(t) \cdot E_s + \left(\sum_{i=1}^{6} r_i\right)\left(E_a - \frac{E_0 \cdot A_T}{10 \cdot \left(\sum_{i=1}^{6} r_i\right)}\right)\right)}{p(t) \cdot E_s \cdot E_c \cdot (1 + H(t))} \end{aligned} \tag{13}$$

### 6.3.3 Additions to reserves

The reserves are augmented by the factor $\frac{H(t)}{1+H(t)}$ of the food collected in that month. There are three cases to be considered,

1. sustenance energy from previous month adequate for collection needs
2. reserves tapped to obtain energy for collection of food
3. reserves inadequate for energy needs

**Adequate energy levels** In situations in which the energy obtained from the previous month was adequate to collect the required number of acorns, the number of acorns placed into storage would be,

$$\begin{aligned} r_0 &= \frac{H(t)}{1 + H(t)} \cdot \frac{p(t) \cdot E_s}{E_a}(1 + H(t)) \\ &= H(t) \cdot \frac{p(t) \cdot E_s}{E_a} \end{aligned} \tag{14}$$

**Reserves utilised** When the reserves provide an adequate amount of energy to fulfill all collection requirements, the number of acorns collected would be the same as in the case of adequate energy levels, thus the contribution to the reserves would be, $H(t) \cdot \frac{p(t) \cdot E_s}{E_a}$.

**Inadequate reserves** In the case of inadequate reserves, the number of acorns collected was less than the required amount, thus we obtain the following expression for the increase in food reserves,

$$\begin{aligned} r_0 &= \frac{H(t)}{1+H(t)} \cdot \frac{E_{total}(t)}{E_c} \\ &= \frac{H(t) \cdot \left(p(t) \cdot E_s + \left(\sum_{i=1}^{6} r_i\right)\left(E_a - \frac{E_0 \cdot A_T}{10 \cdot \left(\sum_{i=1}^{6} r_i\right)}\right)\right)}{(1+H(t)) \cdot E_c} \end{aligned} \tag{15}$$

### 6.3.4 Fitness Function as a measure of Zygotic Selection

Recall in equation (13), we found an expression for the survival factor when the reserves are inadequate to sustain the current population level. This inadequate food supply results in population attrition and consequently measures the intensity of zygotic selection. We might consider the surviving population relative to the initial population to be the product of each month's survival proportion,

$$P_{survival} = \prod_{t=1}^{12} P_{survival}(t) \tag{16}$$

### 6.3.5 Food Reserves and Fecundity Selection

We have defined the reproductive cycle of the squirrel to occur at the end of the yearly cycle, and this serves as the second consideration in determining the fitness function. We shall assume that the fecundity is proportional to the remaining population level and the amount of reserves available. Each new population unit would require $E_{new}$ to produce. Thus, the fitness function has the form,

$$Fitness = P_{survival} \cdot \left(1 + \frac{E_a \cdot \left(\sum_{i=1}^{6} r_i\right)}{E_{new}}\right) \tag{17}$$

### 6.3.6 Normalisation of Fitness Function

While it may seem necessary to normalise the fitness function, it adds an unnecessary level of computation to the simulation, since we are only concerned with the relative adaptive advantage of the genotypes in consideration.

## 6.4 Algorithm for the computation of the Fitness Function

We have described and quantified in the previous section the processes which will affect the squirrel population in a single iteration of a yearly mating cycle. We have therefore developed all the necessary mathematical machinery to analyse the adaptive advantage associated with a particular hedging strategy.

In this section, we shall combine the definitions and processes described in the preceding sections to develop an algorithm to compute the fitness function for a specific food supply function $F(t)$, and a given hedging intensity configuration, $H(t)$.

The pseudo-code for computing the Fitness Function is as follows,

**Function Fitness** ( $F(t)$, $H(t)$ ) : real;

**begin**

{ *Initialisation* }

$t = 1$

$p(t) = 1$

$R(t) = (0, 0, 0, 0, 0, 0)$

**while** $t \leq 12$ **do begin**

{ *Zygotic Selection* }

$$Case(t) = \begin{cases} 1 & p(t) \cdot E_s \geq \frac{p(t) \cdot E_s}{1.5 \cdot F(t)} (1 + H(t)) \\ 2 & p(t) \cdot E_s + \left(\sum_{i=1}^{6} r_i\right) \left(E_a - \frac{E_0 \cdot A_T}{10 \cdot \left(\sum_{i=1}^{6} r_i\right)}\right) \geq \frac{p(t) \cdot E_s}{1.5 \cdot F(t)} (1 + H(t)) \\ 3 & p(t) \cdot E_s + \left(\sum_{i=1}^{6} r_i\right) \left(E_a - \frac{E_0 \cdot A_T}{10 \cdot \left(\sum_{i=1}^{6} r_i\right)}\right) < \frac{p(t) \cdot E_s}{1.5 \cdot F(t)} (1 + H(t)) \end{cases}$$

$$P_{survival}(t) = \begin{cases} 1 & Case(t) = 1, 2 \\ \frac{E_a \left(p(t) \cdot E_s + \left(\sum_{i=1}^{6} r_i\right) \left(E_a - \frac{E_0 \cdot A_T}{10 \cdot \left(\sum_{i=1}^{6} r_i\right)}\right)\right)}{p(t) \cdot E_s \cdot E_c \cdot (1 + H(t))} & Case(t) = 3 \end{cases}$$

$$R_{consumed} = \begin{cases} 0 & Case(t) = 1 \\ \left(\frac{p(t) \cdot \left(\frac{E_s}{1.5 \cdot F(t)} (1 + H(t)) - E_s\right)}{E_a - \frac{E_0 \cdot A_T}{10 \cdot \left(\sum_{i=1}^{6} r_i\right)}}\right) & Case(t) = 2 \\ \left(\sum_{i=1}^{6} r_i\right) & Case(t) = 3 \end{cases}$$

$i = 6$

**while** $R_{consumed} > r_i, i > 0$ **do** $\{R_{consumed} \mapsto R_{consumed} - r_i, i \mapsto i - 1\}$

$$R'(t) = (\underbrace{r_1, \ldots r_{i-1}, r_i - R_{consumed}}_{i \text{ terms}}, \underbrace{0, \ldots 0}_{6-i \text{ terms}})$$

$$r_o = \begin{cases} H(t) \cdot \frac{p(t) \cdot E_s}{E_a} & Case(t) = 1, 2 \\ \frac{H(t) \cdot \left(p(t) \cdot E_s + \left(\sum_{i=1}^{6} r_i\right) \left(E_a - \frac{E_0 \cdot A_T}{10 \cdot \left(\sum_{i=1}^{6} r_i\right)}\right)\right)}{(1 + H(t)) \cdot E_c} & Case(t) = 3 \end{cases}$$

$R(t+1) = (r_0, dr'_1, dr'_2, dr'_3, dr'_4, dr'_5)$

$p(t+1) = p(t) \cdot P_{survival}(t)$

$t = t + 1$

**end;** { *while* $t \leq 12$ }

$$Fitness = \left(\prod_{t=1}^{12} P_{survival}(t)\right) \cdot \left(1 + \frac{E_a \cdot \left(\sum_{i=1}^{6} r_i\right)}{E_{new}}\right) \quad \{ \textit{Zygotic and Fecundity Selection} \}$$

**return** *Fitness*

**end.** { *Fitness* }

### 6.5 Multi-dimensional Optimisation in Genetic Configuration Space

We have defined the computational scheme for evaluating the fitness function, which depends upon a fixed food supply function, $F(t)$. It is thus necessary to probe the genetic configuration space for the parameters $H(t) \in [0, 1]$. The search using *Powell's Method in Multi-dimensions* can be constrained to the required parameter space by defining the Fitness Function exterior to the desired parameter space as being some negative number, thereby bounding the search space to $H(t) \in [0, 1]$.

## 7 Discussion

As stated earlier, the undifferentiated increase in food supply as a consequence of hedging, and the log lag time ∼30 years between the adaptive behaviour and the advantage derived from this behavioural pattern, implies that the effect of natural selection on the basis for this mechanism of hedging resulting in increased food supply, is insignificant compared to mechanisms which would have a more immediate and localised adaptive advantage to the individual exhibiting such an adaptive behaviour.

### 7.1 Periods of Abundance and Scarcity

The formulation of the fitness function presented above is thus explicitly dependent upon the differentiated effect of periods of abundance and scarcity in food supply on individuals with differing food collection strategies. The substantial difference in the inter-generational period of oak trees and squirrels allows us to assume to reasonable accuracy that the food supply function which is dependent upon the genetic configuration of the oak trees in a fashion analogous to the food collection intensity of squirrels, is relatively constant for the period of analysis.

### 7.2 Hedging Strategies manifested in nature

The optimum which is exhibited in nature is primarily dependent upon the food supply function, $F(t)$, which is intimately associated with the food collection strategies which squirrels will adopt. In addition, even if $F(t)$ were to be held constant, it is possible for the topology of the 12-dimensional fitness landscape manifold associated with $F(t)$ in the 13-dimensional embedding space, which is composed of a 12-dimensional genetic configuration space plus a 1-dimensional fitness function coordinate, to affect the strategies which squirrels adopt.

Assuming that the initial gene pool is associated with a set of points in the 12-dimensional genetic configuration space, we might then consider the possible states of the next generation to correspond to all pairwise paths between the points representing the initial genetic configurations.

In reality, the genetic configuration space is discrete, and might be better represented as a lattice structure, but in the limit, we may approximate the configuration space using a continuous treatment.

We might visualise this in the case of a 2-dimensional genetic configuration space with the aid of a graph defined on some lattice of genetic configurations.

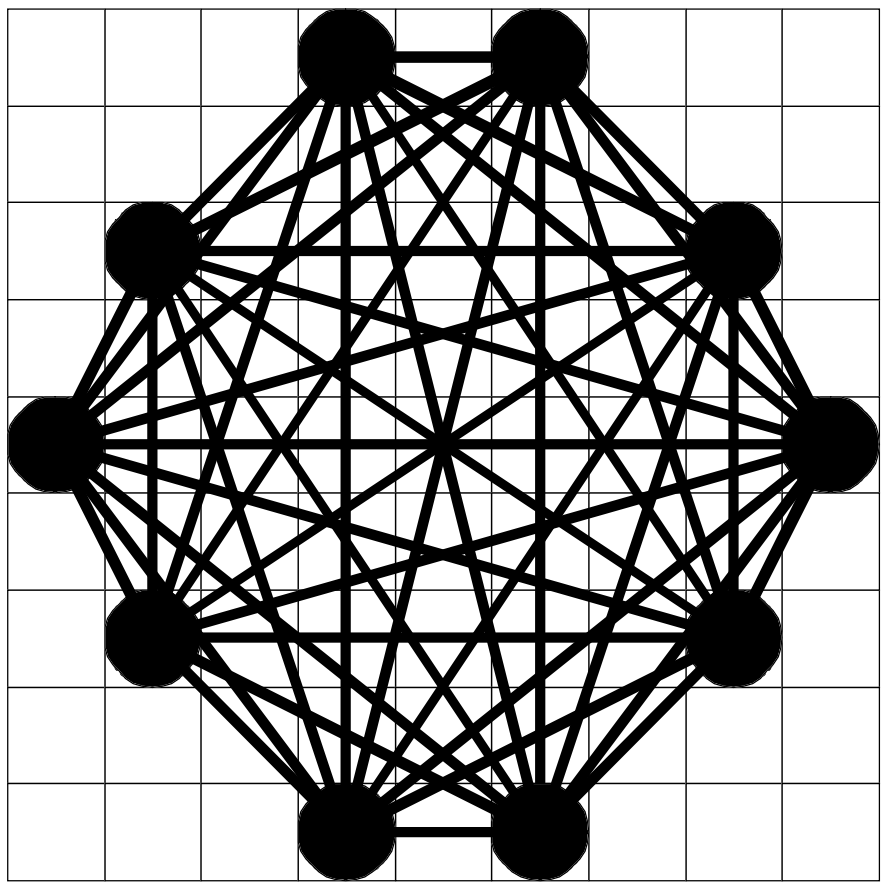

2-dimensional graph analog to allele interaction in genetic space

In the limiting case, with a substantial number of alternative alleles, the process converges onto a greedy algorithm approach to multi-dimensional maximisation.

## 7.3 Analogies between the Negative of the Fitness Function and Potential Functions

We might reasonably consider each successive iteration to cause the convex set of points to diminish in measure, and converge to a local maxima. Therein lies another possibility which could conceivably change the behavioural patterns which will be exhibited in nature.

If we take the negative of the fitness function, the stable points of the iterative procedure might be considered to be analogous to potential minimas. Regions of low adaptive value would then correspond to potential barriers which would constrain the evolution of the system to some potential well. It is for this reason that the stable point which the system eventually converges to depends upon the topology of the fitness landscape in the 13-dimensional embedding space.

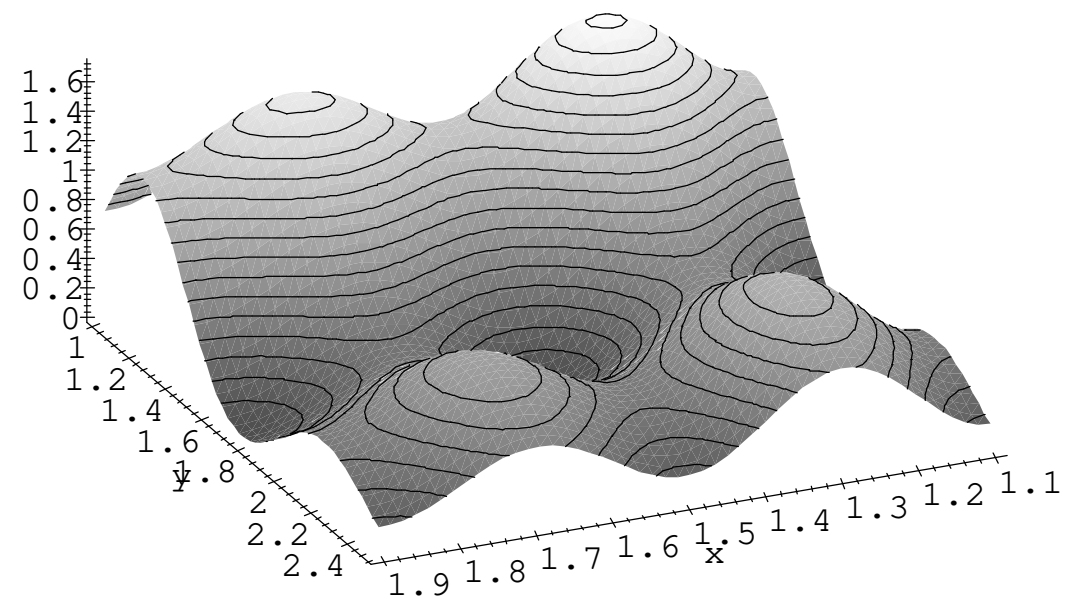

Convergence of evolutionary system to local maximas

Thus, the genetic system may evolve to a local maxima point of the fitness function, which would be locally stable unless substantially perturbed by perhaps a change in the food distribution function. In the fitness landscape illustrated above, it is possible for the system to converge to one of the three local maximas, instead of the global maxima.

### 7.4 Coevolutionary pressure exerted by oak trees and squirrels

Oak trees could conceivably coevolve to vary the amount of acorns provided depending upon the time of the year, since the amount of energy which must be expended for a fixed production would then be inversely related to the energy source, sunlight which has an amplitude envelope with a period of 1 year.

Based upon the assumption that the initial condition for oak trees would be to produce acorns according to a fixed proportion of its energy intake, the production of acorns would be proportional to the amount of sunlight received at the time of the year.

It would be conceivable that squirrels with the propensity to collect reserves of food immediately prior to periods of scarcity would tend to predominate. The predominance of squirrels with this characteristic would allow oak trees to proportionately reduce their production further during the winter months, without unduly diminishing their viability.

As the inter-generational time for squirrels is approximately 1 year, as compared to something like 30 years for oak trees, it immediately follows that the selection pressure exerted by oak trees upon squirrels is substantially greater than that of squirrels upon oak trees. The time scales also imply that the rate of genetic change for oak trees will be significantly less than that of squirrels.

## 8 Conclusion

We have developed the theoretical framework in this paper to analyse the optimal food collection and hedging strategy for squirrels given a particular food supply function. Since the food supply function is dependent upon the local characteristics of the ecological system, the specific analysis of optimal strategies in specific cases will not be considered in this paper.

It would however be fruitful to analyse the food supply and hedging functions of an actual biological system of squirrels and oak trees, and compare the empirical results with the theoretical predictions of this model.